\documentclass[a4paper,12pt]{article}
\usepackage[T2A]{fontenc}
\usepackage[cp1251]{inputenc}  
\usepackage[russian, english]{babel}
\usepackage[dvips]{graphicx}
\usepackage{amsmath}
\usepackage{amssymb}
\usepackage{bm}
\usepackage{indentfirst}

\newcommand{\beq}{\begin{equation}}
\newcommand{\eeq}{\end{equation}}
\newcommand{\beqn}{\begin{eqnarray}}
\newcommand{\eeqn}{\end{eqnarray}}

\newcommand{\al}{\mbox{${\alpha}$}}

\newcommand{\ga}{\mbox{${\gamma}$}}

\newcommand{\de}{\mbox{${\delta}$}}

\newcommand{\ep}{\mbox{${\varepsilon}$}}

\begin{document}

\begin{center}
{\bf \large Electromagnetic production of electron and positron\\
in collisions of heavy nuclei}
\end{center}

\begin{center}
I.B. Khriplovich\footnote{khriplovich@inp.nsk.su}
\end{center}
\begin{center}
Budker Institute of Nuclear Physics\\
630090 Novosibirsk, Russia
\end{center}

\bigskip

\begin{abstract}

We consider the cross-section of electromagnetic production of
$e^+ e^-$ pair in adiabatic scattering of heavy nuclei.

\end{abstract}

\vspace{8mm}

It is well-known that for a point-like nucleus of charge $Z$ the
ground state energy\hspace{8mm}  [1 - 3]
\beq
\ep_0 = m\,\sqrt{1 - Z^2 \al^2}
\eeq
of a single-electron ion turns to zero for $Z = Z_c = 137$; here and below $m$ is the
electron mass, $\al = 1/137$. The account for the final size of
nucleus shifts the critical value to $Z_c \simeq 170$ \cite{ps,
zp}. At higher $Z$, such superheavy nucleus gets unstable with
respect to the decay
\beq
Z \to Z + e^- + e^+,
\eeq
as well as to the decay into single-electron ion $(Z e^-)$ with
subcritical charge $Z-1$ and positron \cite{gr}:
\beq
Z \to (Z e^-) + e^+.
\eeq

Of course, the creation of such superheavy nuclei by itself does
not look realistic. Still, the phenomenon could be observed under
the adiabatic rapprochement of two sufficiently heavy common
nuclei [7, 8]. Calculations performed in Ref. \cite{po} result in
the estimate
\beq
\sigma_P \sim 10^{-25} \;\rm{cm}^2
\eeq
for the cross-section of positron production in collisions of
uranium nuclei. As to the electron produced simultaneously, it is
captured here by one of the nuclei. This effect is essentially
nonperturbative.

\vspace{5 mm}

In fact, the rapprochement of two sufficiently heavy nuclei is
accompanied also by the production of $e^+ e^-$ pairs due to
rather common QED effects. We analyze these effects, and
demonstrate that they are comparable with (4).

So, let us address the diagram presented in Fig. 1. Double lines
therein describe the propagation of heavy nuclei, single and wavy
lines refer to $e^{\pm}$ and photons, respectively. The essential
property of this diagram is that it does not vanish in the limit
of heavy mass of the nuclei, i.e. for $M \to \infty$.

\begin{figure}[h]
\begin{center}
    \includegraphics[width=2cm]{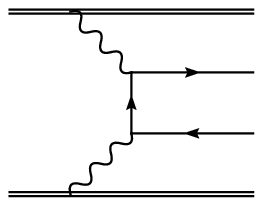}\hskip 1cm
    \includegraphics[width=2cm]{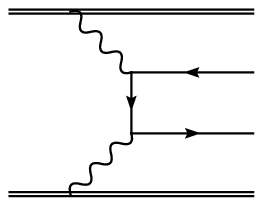}\\
    \vskip 0.6cm
    Fig.1
\end{center}
\end{figure}
The corresponding cross-section looks as follows:
\[
d\sigma = 32\;\frac{Z^4
\al^4}{\pi^4}\;v^3\,\frac{d^3{p}_-}{2p_{-0}}\,
\frac{d^3{p}_+}{2p_{+0}}\, \delta (p_{-0} + p_{+0} - \ep)
\int\frac{d^3{q_1}}{\bar{q_1}^4}
\frac{d^3{q_2}}{\bar{q_2}^4}\,\de(\bar{q}_1+\bar{q}_2+\bar{p}_-
+\bar{p}_+)
\]
\beq \times \; \frac{Sp\,\ga_3(\hat{p}_- + m) \ga_3 (\hat{p}_- -
\hat{q}_1 + m ) \ga_3 (\hat{p}_+ - m) \ga_3 (\hat{p}_+ + \hat{q}_2
+m)}{(\bar{q}_1^2 - 2\bar{q}_1\bar{p}_-) (\bar{q}_2^2 + 2\bar{q}_2
\bar{p}_+)}\;.
\eeq
Here $\bar{p}_-$, $\bar{p}_+$ are the three-dimensional components
of the momenta of produced electron and positron; $p_{-0}$,
$p_{+0}$ are the time components of these momenta; $\ep$ is the
total energy extracted by the created electron - positron pair
from the colliding heavy nuclei. The virtual quanta
$\bar{q}_{1,2}$, emitted by the nuclei, produce then the $e^+ e^-$
pair. In particular, these vector fields enter as
$1/\bar{q}_{1,2}^2$ the amplitude of the discussed process, and as
$1/\bar{q}_{1,2}^4$ its cross-section. As to the Coulomb
distortion of nuclear orbits, it is inessential here since we
consider in fact small-angle scattering of the colliding nuclei.
For the velocity of nuclei, we assume $v \simeq 0.1$; we assume
also $Z \simeq 92$.

To our approximation, the heavy nuclei can be treated classically.
The photons (wavy lines) are produced here by electromagnetic
currents of heavy nuclei, i.e., by $Zev$. One more power of $e$ is
due to the charges of the created electron and positron. This flux
of the colliding nuclei is proportional to $v$, and enters the
result in the denominator. This is the origin of factor $Z^4 \al^4
v^3$ in formula (5).

As to rather tedious spinor structure
\beq
\ga_3(\hat{p}_- + m)
\ga_3 (\hat{p}_- - \hat{q}_1 + m ) \ga_3 (\hat{p}_+ - m) \ga_3
(\hat{p}_+ + \hat{q}_2 +m)
\eeq
in formula (5), with  eight $\ga$-matrices and common momentum
operators $\hat{p}_\pm = \ga_0 p_{\pm 0} - \ga_1 p_{\pm 1}- \ga_2
p_{\pm 2} - \ga_3 p_{\pm 3},$ it corresponds to the situation when
the velocities of nuclei are aligned along the 3 axis. This
structure can be rewritten as
\beq
(\tilde{p}_- + m) (\hat{p}_- - \hat{q}_1 + m ) (\tilde{p}_+ -
m) (\hat{p}_+ + \hat{q}_2 +m),
\eeq
with four $\ga$-matrices and
\beq
\tilde{p}_\pm = \ga_0 \,p_{\pm 0} - \ga_1 \,p_{\pm 1}- \ga_2
\,p_{\pm 2} + \ga_3\, p_{\pm 3}.
\eeq

Let us consider now the integration range for the momenta
$\hat{q}_{1,2}$. The energy carried by each of these momenta is
obviously
\beq
\bar{p}^2/2M - (\bar{p} - \bar{q}_{1,2})^2/2M \simeq
(\bar{p}\cdot\bar{q}_{1,2})/M = \bar{v}\cdot\bar{q}_{1,2}.
\eeq
Then, the total energy squared of the two quanta equals
$(\bar{v}\cdot(\bar{q}_{1}+\bar{q}_{2}))^2$, and their total
momentum squared is $(\bar{q}_{1}-\bar{q}_{2})^2$. The invariant
mass of the two virtual photons, and therefore of the produced
pair $e^+ e^-$, is
\beq
\xi = (\bar{v}\cdot(\bar{q}_{1}+\bar{q}_{2}))^2 -
(\bar{q}_{1}-\bar{q}_{2})^2 =
(\bar{v}\cdot(\bar{p}_{-}+\bar{p}_{+}))^2 -
(\bar{q}_{1}-\bar{q}_{2})^2 \geq 4m^2,
\eeq
i.e., for the real creation of the pair $e^+ e^-$, this invariant
should exceed $4m^2$. To simplify the further analysis, we average
equation (10) over the orientation of the total momentum
$\bar{p}_{-}+\bar{p}_{+}$ of the pair with respect to the velocity
$\bar{v}$ of the nuclei. It results in
\beq
\xi = \frac{1}{3}\,v^2(\bar{p}_{-}+\bar{p}_{+})^2 -
(\bar{q}_{1}-\bar{q}_{2})^2 \geq 4m^2.
\eeq
Thus, the total momentum squared of the $e^+ e^-$ pair is bounded
from below as follows:
\beq
(\bar{p}_{-}+\bar{p}_{+})^2 > (3/v^2)\,4m^2 = 12\,m^2/v^2.
\eeq
In other words, the typical momenta $(3-4)\,m/v$ of thus produced
electrons and positrons are about 15 MeV. With such high momenta,
we can neglect in our estimates not only the Coulomb final state
interaction between the electron and positron, but also the
Coulomb final state interaction of $e^+ e^-$ with the nuclei.

Let us estimate now the total cross section of $e^+ e^-$
production in the collision of heavy ions. Just for dimensional
reasons, this cross section should be proportional to the ratio
$v^2/12m^2$. Then, the total cross section should obviously
contain the factor $32(Z\al)^4v^3/\pi^4$ and, as it follows from
formula (5), factors $(4\pi/2)^2 = 4\pi^2$ from
$d^3{p}_-/2p_{-0}\times d^3{p}_+/2p_{+0}$, and $4\pi$ from
$d^3{q}_1$ (one power of $4\pi$ disappears here, together with
$\de(\bar{q}_1+\bar{q}_2+\bar{p}_- +\bar{p}_+)$). In this way we
arrive at the following estimate for cross-section (5):
\beq
\sigma = 32 (Z\al)^4 (v^3/\pi^4)\,(4\pi/2)^2\,(4\pi)\,(v^2/12 m^2)
= \frac{128}{3\pi}(Z\al)^4 v^5 /m^2 \sim 0.4 \cdot 10^{-25} \;
{\rm cm}^2.
\eeq

This result for the $e^+ e^-$ production is on the same order of
magnitude as the estimate $\sigma_P \sim 10^{-25} \; {\rm cm}^2$
made in \cite{po} for the positron production cross-section, with
the electron captured by one of the nuclei.

\subsection*{Acknowledgements}

I am grateful to Yu.Ts. Oganessian for attracting attention to
this problem, and to V.F. Dmitriev, V.G. Serbo, V.V. Sokolov,
and O.V. Zhirov
for numerous discussions.

The investigation was supported in part by the Russian Ministry
of Science.

\renewcommand{\bibname}

\end{document}